\documentstyle[twoside,fleqn,epsf,espcrc2]{article}

\newcommand{\grpp}{g_{\rho\pi\!\pi}}
\newcommand{\gnnp}{g_{N\!N\!\pi}}
\newcommand{\ra}{\rightarrow}
\newcommand{\eqdef}{\stackrel{\rm def}{=}}

\newcommand{\qb}{\overline{q}}

\newcommand{\AmS}{{\protect\the\textfont2
  A\kern-.1667em\lower.5ex\hbox{M}\kern-.125emS}}

\title{Hadronic coupling constants in lattice QCD}

\author{{R.L.~Altmeyer}\address{DESY Hamburg, Notkestr.~85, 
          D--22603 Hamburg, FRG},
        M.~G\"ockeler$^{\rm b}$, R.~Horsley\address{HLRZ c/o KFA J\"ulich,
          D--52428 J\"ulich, FRG},
        E.~Laermann\address{Fakult\"at f\"ur Physik, Universit\"at
          Bielefeld, D--33615 Bielefeld, FRG},
        G.~Schierholz$^{\rm b}$ and
        P.M.~Zerwas$^{\rm a}$}

\begin{document}

\begin{abstract}
We report on calculations of the had\-ro\-nic coupling constants $\grpp$
and $\gnnp$ based on 
lattice QCD with four flavors of dynamical staggered
fermions. By computing 2--point and 3--point Green's functions 
we have been able to determine these coupling constants
{\it ab initio} from QCD; the results $\grpp = 4.2 \pm 1.8$ and 
$\gnnp = 14.8 \pm 6$ are compatible with the experimental values. 
\end{abstract}

\maketitle

Recent lattice QCD calculations of the had\-ro\-nic mass spectrum
have given encouraging results, reproducing many of the experimental
values even at a quantitative level.
This motivates the launching of more complicated analyses
such as the calculation of
properties of the phenomenological hadronic interactions.  
The aim of our work is a
calculation of coupling constants which parameterize the
strength of the interaction between hadrons at low energies.

We have based the analysis on the staggered fermion action 
with four degenerate flavors of dynamical fermions. The lattice size 
is $16^3 \times 24$ and we have generated so far 85 
configurations at $\beta=5.35$, $m=0.01$ using the 
hybrid Monte Carlo algorithm. Details with respect to the production
of configurations can be recalled from Ref.\cite{us}.\\[2mm]
{\bf Hadronic Couplings in the Continuum}

Interactions between ha\-drons at low en\-er\-gies can 
be des\-cribed by
effective La\-gran\-gi\-ans in which the hadrons enter as the ``elementary
particles'' of the theory. The strength of the interaction is
parameterized by effective coupling constants whose values have been
determined by experiment. In our case
the effective interaction Lagrangians are
\begin{eqnarray}
{\cal L}_{Int}^{eff}(x) & = & \grpp \rho_\nu^a (x) \pi^b (x)
\stackrel{\leftrightarrow}{\partial^\nu} \pi^c (x) c_{abc} \\
{\cal L}_{Int}^{eff}(x) & = & \gnnp \overline{N}^a (x) \gamma_5
N^b(x) \pi^c (x) c'_{abc} \nonumber
\end{eqnarray}
\noindent The Lagrangians are
scalars with respect to Lorentz and flavor
transformations, the latter being achieved by contracting the
flavor indices with the appropriate $SU(N_f)$
coefficients $c_{abc}$ and $c'_{abc}$.

The $\grpp$ coupling can be fixed
by measuring the width of the $\rho$ decaying into two pions
giving the experimental value $\grpp^{exp} \simeq 6.08$.

The coupling constant $\gnnp$ can be extracted from a low energy
phase shift analysis of $\pi N$ elastic scattering,
leading to the value of $\gnnp^{exp} \simeq 13.4$.\\[2mm]
{\bf Hadronic Couplings and 3-point Functions}
 
N-point Greens function are calculated on the lattice by means of the path
integral method; 
the hadronic coupling constants demand the
calculation of 3-point functions.

To give an example, we consider the case of the $\rho\pi\pi$
coupling. Here the following 3-point function must be
computed:
\begin{eqnarray}\label{grppdef}
\lefteqn{\Gamma_3(t_1, t_2)  \eqdef  c_{abc} \epsilon_\mu(\vec{q}_\rho)}
\nonumber \\
& & \langle
{\cal T} \{ \tilde{\rho}^\mu_a (\vec{q}_\rho, t_1) \tilde{\pi}_b
(\vec{q}_\pi, t_2) \pi_c(0) \} \rangle.
\end{eqnarray}
$\epsilon_\mu$ is the polarization vector of the $\rho$ meson and 
the tilde indicates
that the corresponding field operator is defined 
in momentum space. Inserting
complete sets of particle states, using the LSZ formalism and applying the
phenomenological interaction Lagrangian, the 
following relation between $\Gamma_3$ and $\grpp$  can be derived 
$(t_1 \gg t_2 \gg 0)$:
\begin{eqnarray}\label{threegrpp}
\lefteqn{
\Gamma_3(t_1,t_2) \simeq \grpp c_{abc} \frac{\sqrt{Z_{\rho_a} Z_{\pi_b}
Z_{\pi_c}}}{8 E_{\rho_a} E_{\pi_b} E_{\pi_c}} 2 |\vec{q}_\rho|
\frac{m_\pi}{m_\rho} } \nonumber \\
& & \frac{2 E_{\pi_c}}{(E_{\rho_a} - E_{\pi_b})^2
- E_{\pi_c}^2} e^{-i E_{\rho_a}(t_1 - t_2)} e^{-i E_{\pi_b} t_2} 
\end{eqnarray}
Two terms in this formula depend on the 
specific process: The tensor $c_{abc}$ and the
kinematical factor $2|\vec{q}_\rho | \frac{m_\pi}{m_\rho}$. The other
terms are analogous for the calculation of the $NN\pi$ coupling constant. 

To determine the value of $\grpp$ 
the knowledge of the masses $m_i$, the energies
$E_i$ and the renormalization constants $Z_i$ of the particles is
required in addition. These observables
have been evaluated in a previous analysis of the spectrum by
computing the proper 2-point functions \cite{us}.\\[2mm] 
{\bf Hadronic Operators on the Lattice}
 
Different approaches can be applied to the construction of the
appropriate hadron operators from the single-component staggered
fermion fields $\chi$.
The application of
group theoretical methods is an adequate and 
mathematically rigid method to construct hadron operators with
staggered fermion fields. In our case, however, a more 
intuitive way is preferable to simplify the calculation of the group
coefficients.
First, quark spinor operators 
are constructed which are then used to build up the hadron operators.
In this way the flavor of the lattice operators can be defined in 
terms of the usual isospin, strangeness and charm quantum numbers,
allowing us to use standard tables of $SU(4)$ Clebsch--Gordan
coefficients \cite{su4const1,su4const2} to calculate the group tensors.

The quark fields are defined
on a {\it coarse} lattice whose points $y$ consist of hypercubes of the
original lattice and the ``{\it fine} degrees of freedom'' 
are used to provide them with spin and flavor indices:
\begin{equation}
q^{\alpha a}(y) \eqdef \sum\nolimits_\eta \Gamma_\eta^{\alpha a} \chi(2y +
\eta).
\end{equation}
\noindent Here $\Gamma_\eta^{\alpha a} = \left( \gamma_1^{\eta_1}
\cdots \gamma_4^{\eta_4}\right)^{\alpha a}$ with 
$\alpha$, $a$ denoting spin and flavor indices of
the quark, respectively. 

Mesons are of the form
\begin{equation}
M_{SF}(y) = \qb(y) [ \Gamma_S \otimes \Gamma_F ] q(y)
\end{equation}
\noindent where the multi-indexed products of $\gamma$ matrices
$\Gamma_S$ and $\Gamma_F$ 
determine the spin and flavor content of the state. In terms of the
staggered fermion fields $\chi$ these operators are non-local if
the spin multi-index $S$ and the flavor multi-index $F$ are not equal.

The local baryon operator which corresponds to the nucleon in the continuum
limit can be constructed as follows
\begin{eqnarray}
B^{\alpha a}(y) & = & \frac{1}{3!} \epsilon_{ijk} 
\sum_A \left(\Gamma_A q_i(y) \Gamma^\dagger_A\right)^{\alpha a}
\times
\nonumber \\
& & \times \left\{ q_j(y) [{\cal C} \Gamma_A \otimes {\cal C}^* \Gamma^*_A]
q_k(y) \right\}.
\end{eqnarray}
\noindent ${\cal C}$ denotes the charge conjugation operator and a
summation over color indices $i,j,k$ is understood. 
$\alpha$ and $a$ are spin and flavor indices 
of the baryon.\\[2mm] 
{\bf Hadron Masses}

We merely recall here the results of the analysis
for the spectrum of the required particles; for details see \cite{us}.
A few remarks, however, ought to be added:
(i) Some of the results are technically new as the
masses and energies had to be computed 
for polarized $\rho$ mesons and nucleons.
(ii) The particle energies have been determined for states 
with minimal lattice momentum\footnote{$L=16$ is the spatial extent of
the lattice.} $q_{min} = \frac{2\pi}{L}$
in the $z$ direction. Fixing
the average $\rho$ mass to $770$ GeV, this corresponds to $q_{min} \sim
550$ GeV in physical units. 
(iii) For the $\rho$ we used a local-time (LT)  as well as a
non-local-time (NLT) operator, the latter supressing the unwanted parity 
partner of the
$\rho$. In the case of the nucleon we used a wall operator for the source.

\vspace{1mm}

\renewcommand{\arraystretch}{1.2}
\noindent
\begin{center}
\begin{tabular}{|l||l|l|}
\hline
 & $m$ & $E(q)$ \\
\hline
$\pi[2]$ & 0.271(5)  & 0.50(3)  \\
$\pi[7]$ & 0.346(15)  &  0.53(3)  \\
\hline
$\rho^{LT}_{pol}[4]$ & 0.55(5)  & 0.70(8) \\
$\rho^{NLT}_{pol}[4]$ & 0.53(2)  & 0.65(9) \\
\hline
$N_{pol}$ & 0.76(7)  & 0.84(8) \\
\hline
\end{tabular}
\end{center}
\renewcommand{\arraystretch}{1}

\vspace{1mm}

\noindent Table 1: Masses and energies in lattice units. The
numbers in square brackets refer to Ref.\cite{us}.\\[2mm]
{\bf The $\grpp$ Coupling Constant}

We have calculated the 3-point function eq.\ (\ref{grppdef}) with the $\pi$
meson being fixed at the 
time slice $t_\pi = 5$ for all values of $t_\rho$.
The $\rho$ meson and one $\pi$ meson were given a momentum $q_{min}$ in
the $z$ direction. 
To avoid statistical noise, the operators were chosen as local 
as possible. We started with a spatially 
local $\rho$ meson and
the local ``lattice Goldstone''-pion, the latter having
the additional advantage of being the lightest pion and having no
parity partner. However, due to the flavor 
symmetry the $\rho$ cannot decay into 
two local $\pi$'s so that the second pion had to be chosen non-local.
The calculation was performed by using both LT 
and NLT $\rho$ operators.

The results which are achieved by doing $\chi^2$ fits of the data to 
eq.\ (\ref{threegrpp}) are:
$$
\left.
\begin{array}{lcll}
\grpp & = & 3.6 \pm 2.5 & \mbox{(LT)} \\
\grpp & = & 4.7 \pm 2.8 & \mbox{(NLT)} 
\end{array}
\right\} \grpp = 4.2 \pm 1.8
$$
\def\epsfsize#1#2{0.5\textwidth}
\begin{figure}[htb]
\epsfclipon
\vspace{-16mm}
\epsffile{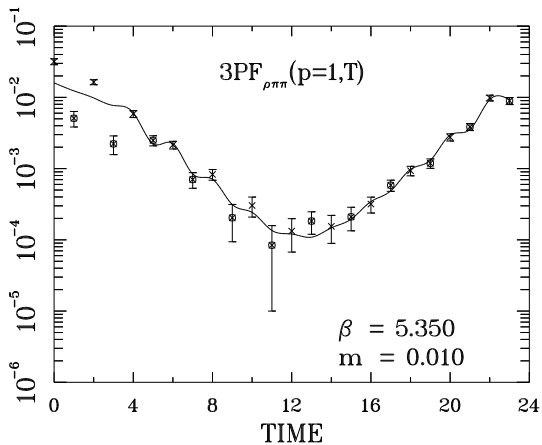}
\vspace{-9mm}
\caption{The data and a $\chi^2$ fit for $\grpp(LT)$}
\vspace{-7mm}
\end{figure}
Though the values are systematically low, within the rather large errors
they are nevertheless 
compatible with the experimental value. Moreover, one should 
bear in mind that only one pion was light in the
$\rho$ decay so that the particles had to be off-shell.\\[2mm]
{\bf The $\gnnp$ Coupling Constant}

We have computed the 3-point function using a polarized nucleon, a local
lattice GS-$\pi$ and a wall-source nucleon operator (only half
of the configurations have been analyzed so far).
The wall
source projects onto a nucleon at rest while the sink nucleon and
pion have been given a momentum $q_{min}$ in the 
$z$ direction.
The sink nucleon has been fixed at the timeslice $t_N = 8$ and we have
evaluated the correlation function for all times $t_\pi$ of the pion.

Unlike the $\rho\pi\pi$ coupling, $\gnnp$ could not be determined 
yet directly from
the data evaluated so far. This is due to the fact that 
there are two different couplings $g_1^A$ and $g_2^A$ for the $N
\ra N \pi$ process in $SU(4)$.
The result obtained from this analysis, is therefore 
a linear superposition  of two
couplings and additional information is needed to disentangle them.
This information can be extracted from the lattice\footnote{Analysis
in progress.};
however, for the time
being we have used the experimentally
known value of $F_A/D_A = 0.59(5)$ \cite{faoverda} to process the
data.

The final result for the $NN\pi$ coupling is given by:
$$
\gnnp  =  14.8 \pm 6 
$$
The central value is much closer to the experimental data than  the result
for $\grpp$; this may be due to the fact that
the $\gnnp$ analysis does not involve 
``non-Goldstone''
pions. However, the error is still large and must be lowered by
increasing the statistics. Nevertheless, this {\it ab initio}
calculation of the $NN\pi$ coupling from QCD is an exciting first step ---
we think.
\begin{figure}[htb]
\epsfclipon
\vspace{-16mm}
\epsffile{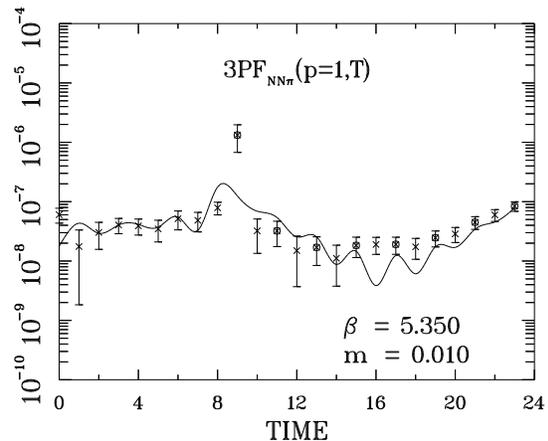}
\vspace{-9mm}
\caption{The data and a $\chi^2$ fit for $\gnnp$}
\vspace{-7mm}
\end{figure}

\end{document}